\documentclass[journal]{IEEEtran} % use the `journal` option for ITherm conference style
\IEEEoverridecommandlockouts
\usepackage{cite}
\usepackage{amsmath,amssymb,amsfonts}
\usepackage{algorithmic}
\usepackage{graphicx}
\usepackage{booktabs}
\usepackage{url}
\usepackage{textcomp}
\usepackage{xcolor}
\def\BibTeX{{\rm B\kern-.05em{\sc i\kern-.025em b}\kern-.08em
    T\kern-.1667em\lower.7ex\hbox{E}\kern-.125emX}}

\usepackage{tikz}
\usetikzlibrary{arrows,decorations,shapes,backgrounds}
\usepackage{pgfplots}
\pgfplotsset{compat=newest}
    
\begin{document}

\title{Low-Power Wake-Up Signal Design in 3GPP Release 18\\
% delete or comment-out the following line before submission
% {\footnotesize \textsuperscript{*}Note: Sub-titles are not captured in Xplore and should not be used}
}

\author{%%%% author names
    \IEEEauthorblockN{1\textsuperscript{st} Sebastian Wagner}% first author
    , \IEEEauthorblockN{2\textsuperscript{nd} Kien Le Trung}% delete this line if not needed
    , \IEEEauthorblockN{3\textsuperscript{rd} Raymond Knopp}% delete this line if not needed
    % duplicate the line above as many times as needed to list all authors
    \\%%%% author affiliations
    \IEEEauthorblockA{\textit{EURECOM, Sophia-Antipolis, France}}\\% first affiliation
    % \IEEEauthorblockA{\textit{dept. name of organization (of Aff.), City, Country if needed}}\\% delete this line if not needed
    % duplicate the line above as many times as needed to list all affiliations
    %%%% corresponding author contact details
    \IEEEauthorblockA{sebastian.wagner@eurecom.fr}
}

\maketitle

\begin{abstract}
    This article provides an overview of the Low-Power Wake-Up Signal (LP-WUS) design in 3GPP Rel-18. A particular focus is the analysis of the different proposed low-power waveform designs in the Rel-18 study item including coding and modulation. The performance of the waveforms is compared through numerical simulations under various channel conditions. Furthermore, a novel coding scheme is proposed that exploits the WUS repetitions in time-domain to transmit additional payload and significantly increases spectral efficiency.
\end{abstract}


\section{Introduction}
Low-Power Wake Up Signals (LP-WUS) are essential in many low power communication protocols such as LoRa, Bluetooth or WiFi \cite{b2}. These signals allow for the design and implementation of low power radios and thus contribute significantly to a reduction in power consumption for devices with application in (Industrial) Internet of Things (IoT). 

In cellular networks, 3GPP has agreed to carry out a Study Item (SI) on LP-WUS \cite{si_wus} in Release 18 with the report available in \cite{b3}. The goal of the SI is to evaluate the potential reduction in power consumption of a 5G device equipped with a LP Wake-Up Radio (WUR). Typically, a 5G device consumes tens of milliwatts even if it is not transmitting or receiving any data. This idle power consumption is due to the fact that the 5G device has to carry out periodic measurements and check for potential paging messages. Hence, 5G-enabled IoT devices without an external power source are difficult to implement in practice. Consequently, the LP-WUR is \textit{independent} of the 5G Main Radio (MR), i.e. the 5G MR can be powered off while the LP-WUR is active and searching for a potential WUS. The LP-WUR functionality may be very similar to existing WUS in LTE and 5G Rel-17 (Paging-Early Indication) which are based on legacy Zadoff-Chu sequences and the Physical Data Control channel (PDCCH), respectively. More precisely, the devices are configured into different groups, each group corresponding to a sequence or a bit position in the PDCCH payload. If a device is in idle mode it will search for the WUS according to the system configuration. If the WUS contains the group identification of the UE, the device will proceed and decode the paging message, otherwise it will go back to sleep and wait for the next WUS occasion.

The LP-WUS SI is divided into three topics, (i) the evaluation of LP-WUS, (ii) LP-WUS receiver architectures and (iii) physical design and procedures of the WUS. The evaluation of LP-WUS focuses on identifying the potential power saving gains, coverage requirements and resource overhead. 

Different receiver architectures are discussed including RF envelope detection, a heterodyne architecture with IF envelope detection, a homodyne/zero-IF architecture with baseband envelope detection as well as various receivers for FSK detection. 

The topic of WUS physical layer design concentrates on issues such as waveform design and comparison, WUS bandwidth and location, measurements, payload content, coding and WUS monitoring procedures.

In this paper, we focus on the WUS physical layer design. In Section \ref{sec:phy-design} a concise introduction of the evaluated waveforms and coding schemes is presented. Subsequently, the receiver design is discussed in Section \ref{sec:rx-design}. Section \ref{sec:td-overlay} proposes a time-domain overlay code to increase spectral efficiency. Numerical results are provided in Section \ref{sec:simulations} and the paper is concluded in Section \ref{sec:conclusion}.

\section{Physical Layer Design}\label{sec:phy-design}

\begin{figure}[htbp]
    \centerline{\includegraphics[width=0.5\textwidth]{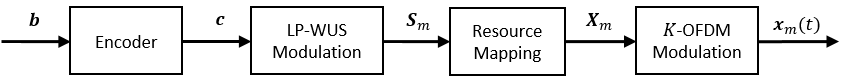}}
    \caption{Block-diagram of LP-WUS waveform generation.}
    \label{fig:wave_generation}
\end{figure}

This section explains how a WUS payload is transmitted on the physical layer. The general requirement is that the WUS generation integrates seamlessly into the legacy NR signal generation. A general block diagram of the WUS transmission is shown in Figure \ref{fig:wave_generation}. The $B$ information bits $\mathbf{b}=[b_0,b_1,...,b_{B-1}]$ are encoded and the resulting $C$ coded bits $\mathbf{c}=[c_0,c_1,...,c_{C-1}]$ are modulated onto $L$ consecutive OFDM symbols each carrying $M$ bits. That is, $M$ is the number of coded bits per OFDM symbol. Subsequently, the WUS in frequency-domain $\mathbf{S}_m$ of message $m=0,1,..,2^{B}-1$ is mapped to the overall resources $\mathbf{X}_m$ of $K$ sub-carriers and OFDM-modulated resulting in the time-domain signal $\mathbf{x}_m(t)$.

Note that in the report \cite{b3}, there is no concise separation between information bits (uncoded bits) and modulated bits. Here, we denote $B$ the total number of \textit{information} bits, $C$ the number of \textit{coded} bits and $M$ the number of modulated bits per OFDM symbol. For instance, if the WUS payload $B=4$ bits is encoded with a Manchester code of rate $R=1/2$, the resulting $C=8$ coded bits are transmitted over $L=4$ OFDM symbols if the WUS modulation scheme supports $M=2$ bits per OFDM symbol.

In general the WUS modulation schemes can be divided into two categories: (i) schemes that use the entire WUS Bandwidth (BW) and (ii) techniques that partition the WUS BW into multiple segments. Each of these categories consists of techniques that can transmit one or more bits per OFDM symbol.

\subsection{Coding}

Prior to WUS modulation, the $B$ information bits $\mathbf{b}$ are encoded to $C$ coded bits $\mathbf{c}$. Denote \textit{codeword} $\mathbf{c}'$ of length $C'\leq C$ encoding bit sequence $\mathbf{b}'$ of length $B'\leq B$. The coding scheme considered by most companies is \textit{Manchester} coding. In its simplest form, a rate $R=1/2$ Manchester code maps a single input bit to two coded bits, i.e.
\begin{align}
    \mathbf{c}' = \begin{cases}
        [0,1] & \text{if } b'=0 \\
        [1,0] & \text{if } b'=1.
    \end{cases}
\end{align}
The main advantage of Manchester coding is that it allows for a simple and robust decoder. More precisely, the decoder simply compares a metric (e.g. received energy) corresponding to the two encoded bits as opposed to a threshold which is difficult to obtain in fading channels. Note that the decoders are discussed in Section \ref{sec:rx-design}.

A straight-forward extension (or generalization) to the above coding scheme can be achieved by jointly encoding more than one bit, such that % $\bar{\mathbf{c}}' = 2^m$
\begin{equation}
\label{eq:e-mc}
    \bar{\mathbf{c}}' = 2^m
\end{equation}
where $\bar{\mathbf{c}}'$ is the \textit{decimal} representation of codeword $\mathbf{c}'$ with length $C'=2^{B'}$ and $m=0,1,...,2^{B'}-1$. The code rate is $R=B'/2^{B'}$, $B'=1$ and $B'=2$ both have code rate $R=1/2$ whereas $B'=3$ yields $R=3/8$.

% It is clear that the position of one-bit in codeword $\mathbf{c}'$ 

\subsection{Multi-Carrier On-Off Keying}

On-Off Keying (OOK) is a well-known modulation allowing for a low-power receiver implementation, i.e. envelope/energy detection. It is a special case of Amplitude Shift Keying (ASK) where there are only two amplitudes, ON and OFF. When applied to a multi-carrier system, such as OFDM, OOK is also referred to as multi-carrier (MC) OOK, because the ON and OFF signals typically span multiple sub-carriers.

Consider the MC-OOK modulated frequency-domain signal $\mathbf{S}_m$ of length $N$ sub-carriers for message $m=0,1,...,2^{B}-1$. Moreover, denote $\mathbf{A}=[A_0,A_1,...,A_{K_S-1}]$ the ON-sequence of length $K_S$ sub-carriers. The OFF-sequence is defined as all zeros.

\begin{figure}[htbp]
    \centerline{\includegraphics[width=0.5\textwidth]{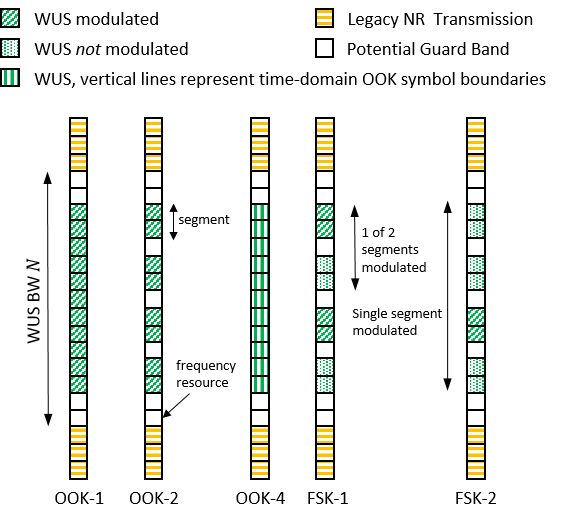}}
    \caption{Resources allocation of LP-WUS waveforms.}
    \label{fig:waveforms}
\end{figure}

The 3GPP LP-WUS SI considers four different OOK schemes and their resource allocation per OFDM symbol is illustrated in Figure \ref{fig:waveforms}.

\subsubsection{OOK-1}
OOK-1 is the classical MC-OOK scheme, where $M=1$ bit is transmitted per OFDM symbol, i.e. $N=K_S$ if no Guard-Band (GB) is considered and $\mathbf{S}_m=\mathbf{A}$ if $c=1$ and $\mathbf{S}_m=0$ if $c=0$.

\subsubsection{OOK-2: Parallel OOK}
A straightforward extension of OOK-1 to $M>1$ is OOK-2, where the available BW $N$ is used to transmit multiple \textit{independent} parallel OOK signals of length $N_M$ sub-carriers, i.e. $\mathbf{S}_m=[\mathbf{S}_0,\mathbf{S}_1,...,\mathbf{S}_{M-1}]$. Evidently, for the same BW $N$, the WUS sequences of the parallel OOK-2 transmissions are significantly shorter than for OOK-1. Moreover, it may be necessary to add a GB between the parallel transmissions to mitigate interference between the transmissions due to the non-ideal filtering at the receiver to extract each individual OOK transmission.
Moreover, since OOK-2 constitutes \textit{independent} OOK transmissions, Manchester coding is applied in \textit{time-domain} which means that the transmit power of the WUS depends on the payload. Consequently, the gNB might not be able to use all the available transmit power for the WUS, sometimes the gNB transmits over the entire WUS BW and at other times only on some segments, which is not ideal.

\subsubsection{OOK-3}
OOK-3 is a scheme where the WUS BW is divided into segments and only a \textit{single} sub-carrier per segment is modulated. The number of parallel OOK transmissions is determined by the length of the segments. The position of the modulated carrier is known to the UE. A special receiver architecture with a Goertzel filter is required to demodulated the signal. We mention this scheme for completeness but it will not be considered in the evaluations.

\subsubsection{OOK-4: Precoded Multi-bit OOK}
\begin{figure}[htbp]
    \centerline{\includegraphics{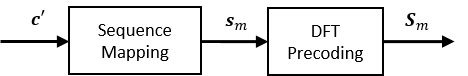}}
    \caption{Block-diagram of OOK-4 modulation per OFDM symbol.}
    \label{fig:ook4-bd}
\end{figure}

A block-diagram of OOK-4 is shown in Figure \ref{fig:ook4-bd}. The coded bits per OFDM symbol $\mathbf{c}'$ (typically $C'=M$) are mapped to \textit{time-domain} sequences $\mathbf{a}$ of length $N_M$ resulting in sequence $\mathbf{s}_m=[\mathbf{s}_{0,m},\mathbf{s}_{1,m},...,\mathbf{s}_{M-1,m}]$ of length $N$, i.e. $\mathbf{s}_{i,m}=\mathbf{a}$ if $c'_i=1$ and $\mathbf{s}_{i,m}=0$ otherwise, $i=0,1,...,M-1$.

\begin{figure}[t]
  \centering
  \begin{tikzpicture}
  \tikzstyle{every pin}=[fill=white,draw=black]
   \pgfplotsset{every axis legend/.append style={
     cells={anchor=west},nodes={scale=0.7, transform shape}}}%, at={(0.5,1.05)}, anchor=south}}
 %   \pgfplotsset{every axis plot/.append style={smooth}}
    \pgfplotsset{every axis/.append style={line width=0.5pt}}
    \pgfplotsset{every axis/.append style={mark options=solid, mark size=2.5pt}}

    \begin{axis}[xlabel={time-domain samples}, ylabel={$\|\mathbf{s}_m\|^2$},
      grid=minor, xmin=1, xmax=256, xtick={1,32,64,...,256}, xmajorgrids,ymajorgrids, ymin=0,
      ymax=0.0004, legend columns=1]

      % OOK-4, M=1, R=1/2
      \addplot[black, densely dashdotted] plot coordinates {(1.000000,0.000060) (2.000000,0.000119) (3.000000,0.000003) (4.000000,0.000090) (5.000000,0.000002) (6.000000,0.000078) (7.000000,0.000013) (8.000000,0.000069) (9.000000,0.000033) (10.000000,0.000061) (11.000000,0.000057) (12.000000,0.000056) (13.000000,0.000084) (14.000000,0.000056) (15.000000,0.000109) (16.000000,0.000063) (17.000000,0.000129) (18.000000,0.000076) (19.000000,0.000142) (20.000000,0.000091) (21.000000,0.000149) (22.000000,0.000104) (23.000000,0.000147) (24.000000,0.000112) (25.000000,0.000138) (26.000000,0.000112) (27.000000,0.000125) (28.000000,0.000105) (29.000000,0.000112) (30.000000,0.000097) (31.000000,0.000104) (32.000000,0.000094) (33.000000,0.000107) (34.000000,0.000100) (35.000000,0.000119) (36.000000,0.000113) (37.000000,0.000136) (38.000000,0.000125) (39.000000,0.000146) (40.000000,0.000127) (41.000000,0.000144) (42.000000,0.000117) (43.000000,0.000133) (44.000000,0.000106) (45.000000,0.000127) (46.000000,0.000106) (47.000000,0.000134) (48.000000,0.000121) (49.000000,0.000146) (50.000000,0.000135) (51.000000,0.000146) (52.000000,0.000136) (53.000000,0.000129) (54.000000,0.000129) (55.000000,0.000116) (56.000000,0.000134) (57.000000,0.000123) (58.000000,0.000149) (59.000000,0.000136) (60.000000,0.000149) (61.000000,0.000135) (62.000000,0.000130) (63.000000,0.000131) (64.000000,0.000121) (65.000000,0.000143) (66.000000,0.000133) (67.000000,0.000152) (68.000000,0.000139) (69.000000,0.000135) (70.000000,0.000135) (71.000000,0.000122) (72.000000,0.000143) (73.000000,0.000136) (74.000000,0.000148) (75.000000,0.000143) (76.000000,0.000129) (77.000000,0.000137) (78.000000,0.000126) (79.000000,0.000143) (80.000000,0.000146) (81.000000,0.000138) (82.000000,0.000143) (83.000000,0.000125) (84.000000,0.000135) (85.000000,0.000143) (86.000000,0.000139) (87.000000,0.000148) (88.000000,0.000130) (89.000000,0.000130) (90.000000,0.000142) (91.000000,0.000137) (92.000000,0.000147) (93.000000,0.000138) (94.000000,0.000126) (95.000000,0.000141) (96.000000,0.000140) (97.000000,0.000141) (98.000000,0.000143) (99.000000,0.000126) (100.000000,0.000135) (101.000000,0.000148) (102.000000,0.000137) (103.000000,0.000137) (104.000000,0.000135) (105.000000,0.000132) (106.000000,0.000147) (107.000000,0.000142) (108.000000,0.000126) (109.000000,0.000137) (110.000000,0.000144) (111.000000,0.000137) (112.000000,0.000138) (113.000000,0.000135) (114.000000,0.000134) (115.000000,0.000147) (116.000000,0.000140) (117.000000,0.000126) (118.000000,0.000140) (119.000000,0.000148) (120.000000,0.000132) (121.000000,0.000131) (122.000000,0.000144) (123.000000,0.000140) (124.000000,0.000133) (125.000000,0.000137) (126.000000,0.000139) (127.000000,0.000137) (128.000000,0.000138) (129.000000,0.000136) (130.000000,0.000136) (131.000000,0.000140) (132.000000,0.000138) (133.000000,0.000134) (134.000000,0.000138) (135.000000,0.000141) (136.000000,0.000135) (137.000000,0.000135) (138.000000,0.000140) (139.000000,0.000138) (140.000000,0.000136) (141.000000,0.000137) (142.000000,0.000137) (143.000000,0.000138) (144.000000,0.000140) (145.000000,0.000134) (146.000000,0.000134) (147.000000,0.000144) (148.000000,0.000139) (149.000000,0.000127) (150.000000,0.000139) (151.000000,0.000148) (152.000000,0.000131) (153.000000,0.000129) (154.000000,0.000146) (155.000000,0.000140) (156.000000,0.000133) (157.000000,0.000137) (158.000000,0.000135) (159.000000,0.000139) (160.000000,0.000145) (161.000000,0.000129) (162.000000,0.000130) (163.000000,0.000148) (164.000000,0.000139) (165.000000,0.000132) (166.000000,0.000137) (167.000000,0.000133) (168.000000,0.000142) (169.000000,0.000146) (170.000000,0.000127) (171.000000,0.000133) (172.000000,0.000142) (173.000000,0.000137) (174.000000,0.000144) (175.000000,0.000132) (176.000000,0.000126) (177.000000,0.000144) (178.000000,0.000139) (179.000000,0.000140) (180.000000,0.000139) (181.000000,0.000123) (182.000000,0.000139) (183.000000,0.000142) (184.000000,0.000138) (185.000000,0.000144) (186.000000,0.000124) (187.000000,0.000132) (188.000000,0.000139) (189.000000,0.000136) (190.000000,0.000150) (191.000000,0.000130) (192.000000,0.000131) (193.000000,0.000132) (194.000000,0.000128) (195.000000,0.000149) (196.000000,0.000136) (197.000000,0.000143) (198.000000,0.000132) (199.000000,0.000123) (200.000000,0.000136) (201.000000,0.000126) (202.000000,0.000149) (203.000000,0.000137) (204.000000,0.000140) (205.000000,0.000133) (206.000000,0.000120) (207.000000,0.000133) (208.000000,0.000120) (209.000000,0.000145) (210.000000,0.000134) (211.000000,0.000145) (212.000000,0.000134) (213.000000,0.000126) (214.000000,0.000128) (215.000000,0.000113) (216.000000,0.000132) (217.000000,0.000118) (218.000000,0.000146) (219.000000,0.000128) (220.000000,0.000147) (221.000000,0.000126) (222.000000,0.000132) (223.000000,0.000116) (224.000000,0.000115) (225.000000,0.000115) (226.000000,0.000111) (227.000000,0.000126) (228.000000,0.000118) (229.000000,0.000141) (230.000000,0.000123) (231.000000,0.000145) (232.000000,0.000117) (233.000000,0.000137) (234.000000,0.000103) (235.000000,0.000123) (236.000000,0.000088) (237.000000,0.000113) (238.000000,0.000081) (239.000000,0.000114) (240.000000,0.000084) (241.000000,0.000123) (242.000000,0.000092) (243.000000,0.000135) (244.000000,0.000100) (245.000000,0.000144) (246.000000,0.000105) (247.000000,0.000144) (248.000000,0.000107) (249.000000,0.000133) (250.000000,0.000105) (251.000000,0.000113) (252.000000,0.000103) (253.000000,0.000089) (254.000000,0.000101) (255.000000,0.000064) (256.000000,0.000098) };

      % OOK-4, M=2, R=1/2
      \addplot[red, dashed] plot coordinates {(1.000000,0.000108) (2.000000,0.000295) (3.000000,0.000014) (4.000000,0.000202) (5.000000,0.000032) (6.000000,0.000180) (7.000000,0.000062) (8.000000,0.000186) (9.000000,0.000103) (10.000000,0.000206) (11.000000,0.000153) (12.000000,0.000234) (13.000000,0.000203) (14.000000,0.000262) (15.000000,0.000242) (16.000000,0.000279) (17.000000,0.000258) (18.000000,0.000277) (19.000000,0.000247) (20.000000,0.000260) (21.000000,0.000222) (22.000000,0.000248) (23.000000,0.000212) (24.000000,0.000262) (25.000000,0.000233) (26.000000,0.000295) (27.000000,0.000265) (28.000000,0.000308) (29.000000,0.000271) (30.000000,0.000278) (31.000000,0.000260) (32.000000,0.000248) (33.000000,0.000276) (34.000000,0.000261) (35.000000,0.000306) (36.000000,0.000280) (37.000000,0.000285) (38.000000,0.000272) (39.000000,0.000251) (40.000000,0.000285) (41.000000,0.000274) (42.000000,0.000295) (43.000000,0.000288) (44.000000,0.000256) (45.000000,0.000276) (46.000000,0.000272) (47.000000,0.000283) (48.000000,0.000296) (49.000000,0.000261) (50.000000,0.000266) (51.000000,0.000284) (52.000000,0.000276) (53.000000,0.000286) (54.000000,0.000273) (55.000000,0.000258) (56.000000,0.000287) (57.000000,0.000290) (58.000000,0.000260) (59.000000,0.000269) (60.000000,0.000288) (61.000000,0.000276) (62.000000,0.000268) (63.000000,0.000276) (64.000000,0.000277) (65.000000,0.000274) (66.000000,0.000274) (67.000000,0.000275) (68.000000,0.000275) (69.000000,0.000274) (70.000000,0.000274) (71.000000,0.000278) (72.000000,0.000271) (73.000000,0.000268) (74.000000,0.000283) (75.000000,0.000280) (76.000000,0.000258) (77.000000,0.000274) (78.000000,0.000293) (79.000000,0.000266) (80.000000,0.000261) (81.000000,0.000281) (82.000000,0.000275) (83.000000,0.000277) (84.000000,0.000276) (85.000000,0.000255) (86.000000,0.000277) (87.000000,0.000288) (88.000000,0.000267) (89.000000,0.000274) (90.000000,0.000260) (91.000000,0.000263) (92.000000,0.000291) (93.000000,0.000270) (94.000000,0.000276) (95.000000,0.000262) (96.000000,0.000250) (97.000000,0.000279) (98.000000,0.000264) (99.000000,0.000288) (100.000000,0.000270) (101.000000,0.000257) (102.000000,0.000261) (103.000000,0.000237) (104.000000,0.000272) (105.000000,0.000252) (106.000000,0.000285) (107.000000,0.000263) (108.000000,0.000267) (109.000000,0.000250) (110.000000,0.000232) (111.000000,0.000238) (112.000000,0.000212) (113.000000,0.000247) (114.000000,0.000216) (115.000000,0.000268) (116.000000,0.000224) (117.000000,0.000282) (118.000000,0.000217) (119.000000,0.000275) (120.000000,0.000191) (121.000000,0.000249) (122.000000,0.000153) (123.000000,0.000216) (124.000000,0.000114) (125.000000,0.000186) (126.000000,0.000082) (127.000000,0.000176) (128.000000,0.000063) (129.000000,0.000316) (130.000000,0.000039) (131.000000,0.000041) (132.000000,0.000004) (133.000000,0.000016) (134.000000,0.000000) (135.000000,0.000008) (136.000000,0.000001) (137.000000,0.000003) (138.000000,0.000003) (139.000000,0.000001) (140.000000,0.000003) (141.000000,0.000000) (142.000000,0.000003) (143.000000,0.000000) (144.000000,0.000002) (145.000000,0.000001) (146.000000,0.000001) (147.000000,0.000001) (148.000000,0.000000) (149.000000,0.000001) (150.000000,0.000000) (151.000000,0.000001) (152.000000,0.000000) (153.000000,0.000001) (154.000000,0.000001) (155.000000,0.000000) (156.000000,0.000001) (157.000000,0.000000) (158.000000,0.000001) (159.000000,0.000000) (160.000000,0.000001) (161.000000,0.000000) (162.000000,0.000000) (163.000000,0.000001) (164.000000,0.000000) (165.000000,0.000001) (166.000000,0.000000) (167.000000,0.000001) (168.000000,0.000000) (169.000000,0.000000) (170.000000,0.000000) (171.000000,0.000000) (172.000000,0.000001) (173.000000,0.000000) (174.000000,0.000001) (175.000000,0.000000) (176.000000,0.000000) (177.000000,0.000000) (178.000000,0.000000) (179.000000,0.000000) (180.000000,0.000000) (181.000000,0.000000) (182.000000,0.000000) (183.000000,0.000000) (184.000000,0.000000) (185.000000,0.000000) (186.000000,0.000000) (187.000000,0.000000) (188.000000,0.000000) (189.000000,0.000000) (190.000000,0.000000) (191.000000,0.000000) (192.000000,0.000000) (193.000000,0.000000) (194.000000,0.000000) (195.000000,0.000000) (196.000000,0.000000) (197.000000,0.000001) (198.000000,0.000000) (199.000000,0.000000) (200.000000,0.000000) (201.000000,0.000000) (202.000000,0.000000) (203.000000,0.000000) (204.000000,0.000001) (205.000000,0.000000) (206.000000,0.000001) (207.000000,0.000000) (208.000000,0.000000) (209.000000,0.000000) (210.000000,0.000000) (211.000000,0.000001) (212.000000,0.000000) (213.000000,0.000001) (214.000000,0.000000) (215.000000,0.000001) (216.000000,0.000000) (217.000000,0.000000) (218.000000,0.000001) (219.000000,0.000000) (220.000000,0.000001) (221.000000,0.000000) (222.000000,0.000001) (223.000000,0.000000) (224.000000,0.000001) (225.000000,0.000001) (226.000000,0.000000) (227.000000,0.000001) (228.000000,0.000000) (229.000000,0.000001) (230.000000,0.000000) (231.000000,0.000001) (232.000000,0.000000) (233.000000,0.000001) (234.000000,0.000001) (235.000000,0.000000) (236.000000,0.000002) (237.000000,0.000000) (238.000000,0.000002) (239.000000,0.000000) (240.000000,0.000002) (241.000000,0.000001) (242.000000,0.000001) (243.000000,0.000003) (244.000000,0.000000) (245.000000,0.000004) (246.000000,0.000000) (247.000000,0.000004) (248.000000,0.000002) (249.000000,0.000003) (250.000000,0.000006) (251.000000,0.000001) (252.000000,0.000012) (253.000000,0.000000) (254.000000,0.000018) (255.000000,0.000008) (256.000000,0.000023) };

      % OOK-4, M=4, R=1/2
      \addplot[blue, solid] plot coordinates {(1.000000,0.000092) (2.000000,0.000278) (3.000000,0.000018) (4.000000,0.000203) (5.000000,0.000052) (6.000000,0.000208) (7.000000,0.000104) (8.000000,0.000250) (9.000000,0.000172) (10.000000,0.000305) (11.000000,0.000237) (12.000000,0.000337) (13.000000,0.000266) (14.000000,0.000317) (15.000000,0.000255) (16.000000,0.000263) (17.000000,0.000256) (18.000000,0.000250) (19.000000,0.000306) (20.000000,0.000280) (21.000000,0.000316) (22.000000,0.000274) (23.000000,0.000256) (24.000000,0.000281) (25.000000,0.000270) (26.000000,0.000300) (27.000000,0.000290) (28.000000,0.000249) (29.000000,0.000275) (30.000000,0.000295) (31.000000,0.000272) (32.000000,0.000269) (33.000000,0.000277) (34.000000,0.000276) (35.000000,0.000275) (36.000000,0.000273) (37.000000,0.000276) (38.000000,0.000274) (39.000000,0.000265) (40.000000,0.000279) (41.000000,0.000286) (42.000000,0.000254) (43.000000,0.000263) (44.000000,0.000294) (45.000000,0.000269) (46.000000,0.000266) (47.000000,0.000266) (48.000000,0.000248) (49.000000,0.000288) (50.000000,0.000274) (51.000000,0.000267) (52.000000,0.000267) (53.000000,0.000225) (54.000000,0.000263) (55.000000,0.000231) (56.000000,0.000278) (57.000000,0.000255) (58.000000,0.000270) (59.000000,0.000246) (60.000000,0.000224) (61.000000,0.000207) (62.000000,0.000161) (63.000000,0.000174) (64.000000,0.000111) (65.000000,0.000279) (66.000000,0.000029) (67.000000,0.000029) (68.000000,0.000003) (69.000000,0.000010) (70.000000,0.000000) (71.000000,0.000005) (72.000000,0.000001) (73.000000,0.000002) (74.000000,0.000002) (75.000000,0.000000) (76.000000,0.000002) (77.000000,0.000000) (78.000000,0.000002) (79.000000,0.000000) (80.000000,0.000001) (81.000000,0.000001) (82.000000,0.000000) (83.000000,0.000001) (84.000000,0.000000) (85.000000,0.000001) (86.000000,0.000000) (87.000000,0.000001) (88.000000,0.000000) (89.000000,0.000001) (90.000000,0.000001) (91.000000,0.000000) (92.000000,0.000001) (93.000000,0.000000) (94.000000,0.000001) (95.000000,0.000000) (96.000000,0.000001) (97.000000,0.000001) (98.000000,0.000000) (99.000000,0.000001) (100.000000,0.000000) (101.000000,0.000001) (102.000000,0.000000) (103.000000,0.000001) (104.000000,0.000000) (105.000000,0.000001) (106.000000,0.000001) (107.000000,0.000000) (108.000000,0.000001) (109.000000,0.000000) (110.000000,0.000001) (111.000000,0.000000) (112.000000,0.000001) (113.000000,0.000001) (114.000000,0.000001) (115.000000,0.000002) (116.000000,0.000000) (117.000000,0.000003) (118.000000,0.000000) (119.000000,0.000003) (120.000000,0.000001) (121.000000,0.000002) (122.000000,0.000004) (123.000000,0.000001) (124.000000,0.000008) (125.000000,0.000000) (126.000000,0.000013) (127.000000,0.000006) (128.000000,0.000017) (129.000000,0.000092) (130.000000,0.000278) (131.000000,0.000018) (132.000000,0.000203) (133.000000,0.000052) (134.000000,0.000208) (135.000000,0.000104) (136.000000,0.000250) (137.000000,0.000172) (138.000000,0.000305) (139.000000,0.000237) (140.000000,0.000337) (141.000000,0.000266) (142.000000,0.000317) (143.000000,0.000255) (144.000000,0.000263) (145.000000,0.000256) (146.000000,0.000250) (147.000000,0.000306) (148.000000,0.000280) (149.000000,0.000316) (150.000000,0.000274) (151.000000,0.000256) (152.000000,0.000281) (153.000000,0.000270) (154.000000,0.000300) (155.000000,0.000290) (156.000000,0.000249) (157.000000,0.000275) (158.000000,0.000295) (159.000000,0.000272) (160.000000,0.000269) (161.000000,0.000277) (162.000000,0.000276) (163.000000,0.000275) (164.000000,0.000273) (165.000000,0.000276) (166.000000,0.000274) (167.000000,0.000265) (168.000000,0.000279) (169.000000,0.000286) (170.000000,0.000254) (171.000000,0.000263) (172.000000,0.000294) (173.000000,0.000269) (174.000000,0.000266) (175.000000,0.000266) (176.000000,0.000248) (177.000000,0.000288) (178.000000,0.000274) (179.000000,0.000267) (180.000000,0.000267) (181.000000,0.000225) (182.000000,0.000263) (183.000000,0.000231) (184.000000,0.000278) (185.000000,0.000255) (186.000000,0.000270) (187.000000,0.000246) (188.000000,0.000224) (189.000000,0.000207) (190.000000,0.000161) (191.000000,0.000174) (192.000000,0.000111) (193.000000,0.000279) (194.000000,0.000029) (195.000000,0.000029) (196.000000,0.000003) (197.000000,0.000010) (198.000000,0.000000) (199.000000,0.000005) (200.000000,0.000001) (201.000000,0.000002) (202.000000,0.000002) (203.000000,0.000000) (204.000000,0.000002) (205.000000,0.000000) (206.000000,0.000002) (207.000000,0.000000) (208.000000,0.000001) (209.000000,0.000001) (210.000000,0.000000) (211.000000,0.000001) (212.000000,0.000000) (213.000000,0.000001) (214.000000,0.000000) (215.000000,0.000001) (216.000000,0.000000) (217.000000,0.000001) (218.000000,0.000001) (219.000000,0.000000) (220.000000,0.000001) (221.000000,0.000000) (222.000000,0.000001) (223.000000,0.000000) (224.000000,0.000001) (225.000000,0.000001) (226.000000,0.000000) (227.000000,0.000001) (228.000000,0.000000) (229.000000,0.000001) (230.000000,0.000000) (231.000000,0.000001) (232.000000,0.000000) (233.000000,0.000001) (234.000000,0.000001) (235.000000,0.000000) (236.000000,0.000001) (237.000000,0.000000) (238.000000,0.000001) (239.000000,0.000000) (240.000000,0.000001) (241.000000,0.000001) (242.000000,0.000001) (243.000000,0.000002) (244.000000,0.000000) (245.000000,0.000003) (246.000000,0.000000) (247.000000,0.000003) (248.000000,0.000001) (249.000000,0.000002) (250.000000,0.000004) (251.000000,0.000001) (252.000000,0.000008) (253.000000,0.000000) (254.000000,0.000013) (255.000000,0.000006) (256.000000,0.000017)};

      \legend{ {$M=1$}\\
               {$M=2$}\\
               {$M=4$}\\};

    \end{axis}
  \end{tikzpicture}
  \caption{Ideal received OOK-4 time-domain waveform after LP-filtering and down-sampling, $\mathbf{b}'=\{\mathbf{1}\}$ with Manchester coding.}
  \label{fig:ook4-rx-samples}
\end{figure}
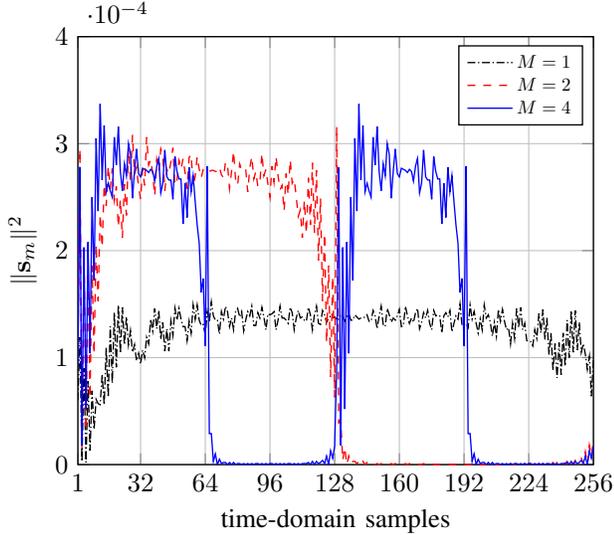

Subsequently, the time-domain signal $\mathbf{s}_m$ is transformed into frequency domain, via DFT-precoding (or Least-squares approximation), before being mapped onto the OFDM resource grid. Prior to DFT-precoding, $\mathbf{s}_m$ may be adapted through pulse-shaping or other signal modification procedures. Similarly, the signal $\mathbf{S}_m$ after DFT-precoding may be modified to alter the spectral shape. For instance, a quantization of the complex values in $\mathbf{S}_m$ to existing QAM symbols can be applied which may reduce the implementation complexity at the gNB, \cite{ericsson}. An example of the ideal, i.e. without noise, received OOK-4 waveform after LP filtering and down-sampling, cf. Table \ref{tab:sim_assumptions}, for different $M$ is shown in Figure \ref{fig:ook4-rx-samples}.

\subsection{Frequency Shift Keying}
Two FSK schemes are considered in the SI, transmitting $B'$ \textit{information} bits ($M>=2$ coded bits) per OFDM symbol, see Figure \ref{fig:waveforms}. Note, that in these FSK schemes, coding and modulation are not readily separated.

\subsubsection{FSK-1}
This scheme divides the WUS BW into $B'$ pairs of non-overlapping segments of length $N_M$, i.e. $B'=M/2$. Consider the corresponding WUS for message $m$, $\mathbf{S}_m=[\mathbf{S}_0,\mathbf{S}_1,...,\mathbf{S}_{2B'-1}]$. Each pair $\{\mathbf{S}_i,\mathbf{S}_{i+1}\}$, $i=0,2,..,2B'-1$ is modulated as
\begin{align}
    \{\mathbf{S}_i,\mathbf{S}_{i+1}\} = \begin{cases}
        \{0, \mathbf{A}\} & \text{if } c=0 \\
        \{\mathbf{A}, 0\} & \text{if } c=1.
    \end{cases}
\end{align}
As in OOK-2, there may be potential GBs between the segments to mitigate inter-segment interference at the receiver. Note that FSK-1 is very similar to OOK-2. In fact, FSK-1 is identical to OOK-2 with Manchester coding $R=1/2$ in frequency-domain. FSK-1 can also be viewed as a frequency-domain equivalent of Manchester-coded ($R=1/2$) OOK-4.

\subsubsection{FSK-2}
FSK-2 divides the WUS BW into $2^{B'}$ non-overlapping segments where only a \textit{single} segment is modulated. Hence, the position of the active segment signals the bits $\mathbf{b}'$. For $B'=1$ this scheme is identical to FSK-1.

Note that FSK-2 and OOK-4 with \textit{extended} Manchester coding in \eqref{eq:e-mc} are equivalent schemes with OOK symbols in frequency and time-domain, respectively. In fact, their performance in ideal conditions is identical. However, FSK-2 is sensitive to the frequency-selectivity of the channel because of the BW partitioning. Whereas OOK-4 is sensitive to timing errors, due to the shorter OOK symbol duration, \cite{b3}.

\subsubsection{Discussion}
All of the above waveforms have their advantages and disadvantages. The robustness of the waveforms to various impairments including frequency-error, timing-error, interference and accuracy of the analog-digital converter (ADC) are evaluated in \cite{b3}. The results suggest that single frequency segment waveforms (e.g. OOK-1, OOK-4) are robust to frequency offsets than multi-frequency segments waveforms (e.g. OOK-2, FSK). Moreover, shortening the symbol duration, e.g. increasing $M$ in OOK-4 or increasing the sub-carrier spacing, increases sensitivity to timing errors. An ADC resolution of 4-bit has been shown to be sufficient for all schemes. Note, that all evaluations assume Manchester coding, without Manchester coding many of the results will not hold true anymore.

The OOK waveforms OOK-1, OOK-3 and OOK-4 where the symbols are distributed over the entire BW are more robust to frequency-selective channels than schemes that use only part of the BW for each bit (OOK-2, FSK). The multi-bit schemes OOK-4 and FSK, increasing $M$, OOK-4 becomes more sensitive to timing errors but is robust to fading, whereas FSK becomes more sensitive to fading but is robust to timing errors. FSK will also entail a more complex receiver design compared to OOK, because multiple receive branches are required to extract the frequency segments and to process the signal. That is why most of the low-power receiver designs favor OOK waveforms \cite{david}.

\subsection{WUS Sequence Design}
In general, the WUS sequence $\mathbf{A}$ in time or frequency domain can be any appropriate sequence, e.g. Gold or Zadoff-Chu sequences. For instance, in 802.11ba \cite{b2} the sequence is composed of any QAM symbols of the specified constellations. Multiple known sequences can be used to encode additional information, but this will require a more complex receiver that is able to carry out correlations. In addition, a known sequence could allow for the design of better receive filters to match the spectral shape.

\section{Receiver Design}\label{sec:rx-design}

The LP-WUS is designed to allow for a low power receiver implementation. A non-coherent envelope detector (ED) can be implemented with very low power consumption and complexity because it does not require phase information of the received signal (no power-hungry PLL required). An ED has only access to the magnitude $|r(t)|$ of the received signal $r(t)$. Consider the case of the AWGN base-band model
\begin{equation}
    r(t) = x_m(t) + n(t)
\end{equation}
with OOK, where $x_0(t)=0$, $x_1(t)=A$ and $n(t)$ is i.i.d. zero-mean Gaussian noise with variance $\sigma^2$. From \cite[Equation 7-4-6]{b1} we obtain the decision rule where the ED decides $x_1$ if
\begin{equation}
    |r| > b ~\text{with}~b=\sigma\sqrt{1+\frac{\gamma}{4}}
\end{equation}
and $x_0$ otherwise, with SNR $\gamma=A^2/\sigma^2$. For fading-channels, computing the threshold $b$ is difficult but can be estimated via a known preamble sequence or computed heuristically and then stored in the receiver. Manchester coding can be utilized to avoid complex threshold determination and improve the detection process. The base-band received signal $r(t)$ is obtained by down-conversion, analog-digital conversion and low-pass (LP) filtering of the WUS. Subsequently, the energy $u_{in}$ corresponding to bit $i$ of codeword $n$ is accumulated over $T$ samples at offset $t_i$, i.e.
\begin{equation}
    u_{in} = \sum_{t=t_i}^{t_i+T}|r(t)|^2.
\end{equation}
Denote $\mathbf{u}_n=[u_{0n},u_{1n},...,u_{C'-1n}]$ the accumulated energy values of each coded bit $c'_i$, $i=0,1,...,C'-1$ in codeword $\mathbf{c}_n'$. The corresponding estimated input message $\hat{m}_n$ reads
\begin{equation}
    \hat{m}_n = \underset{i=0,1,...,C'-1}{\arg\max}\{u_{in}\}.
\end{equation}
The estimated bits $\hat{b}'_n$ are obtained by converting $\hat{m}_n$ into its binary representation.

\section{Time-Domain Overlay Code}\label{sec:td-overlay}
A simple technique to increase the coverage of the LP-WUS is repetition in time-domain, i.e. the WUS $\mathbf{S}_m$ is repeated $L_R$ times. To increase spectral efficiency, it is proposed to transmit additional information $B_h$ by overlaying a signal across the repetitions \cite{knopp1}. This scheme can be applied to any of the WUS waveforms discussed in the previous section, i.e. the payload $B=B_v + B_h$ is split into $B_v$ bits that are encoded and modulated as in Figure \ref{fig:wave_generation} and the $B_h$ bits are transmitted via the time-domain overlay code across the $L_R$ repetitions.

\subsection{Code Generation}
In every OFDM symbol $l$ within a repetition $r$, the WUS is multiplied by a complex symbol $w_{r,m_h}$ depending on message $m_h=0,1,...,2^{B_h}-1$. Denote $\mathbf{w}_{m_h} = [w_{0,m_h},w_{1,m_h},...,w_{L_R-1,m_h}]$ the modulated codeword of length $L_R$ in codebook $\mathcal{W}$. If $B_h\leq L_R$, an orthogonal encoding scheme can be used, e.g. $\mathcal{W}$ is the DFT matrix of size $L_R$ and the modulated codewords $\mathbf{w}_{m_h}$ are the rows or columns of $\mathcal{W}$.
On the other hand, if $B_h > L_R$ we propose to utilize a non-coherent linear code to generate $\mathcal{W}$. For a linear code, the codewords $\mathbf{c}$ are generated from the input bits $\mathbf{b}$ as $\mathbf{c}=\mathbf{bG}$, where $\mathbf{G}$ is the generator matrix of size $B_h\times L_R$.

In order to limit the complexity of the decoder at the LP-WUR, we consider BPSK modulation, i.e.
\begin{equation}
    w_{r,m_h} = e^{-j\pi c_{r,m_h}}, ~ r=0,1,...,L_R-1
\end{equation}
with coded bits $c_{r,m_h}=\{0,1\}$.

A suitable generator matrix $\mathbf{G}=[\mathbf{I}_{B_h} | \mathbf{P}_{B_h\times (L_R-B_h)}]$ can be found, for instance, by the methods described in \cite{knopp_code}. An example used in the numerical evaluations for $B_h=5$ and $L_R=14$ is given by

\begin{equation}
% \footnotesize
\mathbf{P}_{5\times 9} = 
  \begin{bmatrix}
   0 & 1 & 0 & 0 & 1 & 1 & 0 & 0 & 1 \\
   1 & 0 & 1 & 0 & 1 & 1 & 0 & 0 & 1 \\
   1 & 1 & 0 & 0 & 0 & 0 & 1 & 0 & 1 \\
   0 & 0 & 1 & 1 & 0 & 0 & 1 & 0 & 1 \\
   1 & 0 & 0 & 1 & 1 & 0 & 1 & 0 & 1
\end{bmatrix}  
\end{equation}

When applied to the time-domain sequence $\mathbf{s}_{r,m_v}$, the final sequence $\mathbf{s}_{r,m}$ of the combined message $m=m_v+m_h 2^{B_v}$ is obtained as 
\begin{equation}
    \mathbf{s}_{r,m} = \mathbf{s}_{r,m_v} w_{r,m_h}.
\end{equation}

\subsection{Receiver}
An optimal receiver estimates the payload $B=B_v+B_h$, where the $B_v$ bits are transmitted with the LP-WUS waveform, by correlating the received signal $\mathbf{r}_{r,p}$ of repetition $r$ and receive antenna $p$ with all possible messages $\mathbf{s}_{r,m}$, i.e.
\begin{equation}
    \hat{m} = \underset{m}{\arg\max}\left\{\sum_{p=0}^{P-1}\left\| \sum_{r=0}^{L_R-1} \mathbf{r}_{r,p}^H\mathbf{s}_{r,m}\right\|^2\right\}
\end{equation}
where $P$ denotes the number of receive antennas. To reduce the receiver complexity, i.e. the number of correlations, it is proposed to decode $B_v$ \textit{independently} from $B_h$ via energy detection, cf. Section \ref{sec:rx-design}. 
The receiver decodes the $B_h$ bits through correlations with all possible input signal $\mathbf{s}_{r,m_h}$ for a \textit{given} hypothesis $\hat{m}_v$, i.e.
\begin{equation}
    \hat{m}_h = \underset{m_h}{\arg\max}\left\{\sum_{p=0}^{P-1}\left\| \sum_{r=0}^{L_R-1} \mathbf{r}_{r,p}^H\mathbf{s}_{r,m_h,\hat{m}_v}\right\|^2\right\}
\end{equation}
where $\mathbf{s}_{r,m_h,\hat{m}_v}$ are the hypothesis for message $m_h$ given the hypothesis $\hat{m}_v$. 

Compared to a simple ED, decoding the $B_h$ bits requires additional complexity at the WUR. More precisely, a coherent reception is required, i.e. phase information has to be obtained via a power-hungry PLL circuit. Additional processing has to be carried out to compute the correlations with all possible codewords, e.g. 32 correlations for $B_h=5$. Hence, the proposed waveform design allows for the coexistence of WURs with varying degrees of complexity. Low-complexity devices use ED to detect $B_v$ and devices with less stringent power requirements, e.g. devices that have access to an independent power source such as solar, heat, etc., are able decoded $B_h$ \textit{additional} bits.

\section{Simulation Results}\label{sec:simulations}

In this section, we compare the performance in terms of Block Error Rate (BLER) of the various schemes under different channel conditions and averaged over 100k realizations. The general simulation assumptions are summarized in Table \ref{tab:sim_assumptions}. To allow for a fair comparison, we fix the transmit power per OFDM symbol and ensure that the total transmit power over all OFDM symbols is identical.

\begin{table}[ht!]
  \centering
  \begin{tabular}{ll}
    \toprule
    \textbf{Parameter} & \textbf{Value} \\
    \midrule
    \textbf{Carrier Frequency} & 2.6 GHz \\
    \textbf{Sub-carrier Spacing} & 30 kHz \\
    \textbf{System BW} & 20 MHz (51 PRB) \\
    \textbf{Antenna Config} & SISO \\
    \textbf{WUS BW} & 5 MHz (12 PRB) \\
    \textbf{WUS Sampling Rate} & 7.68 MHz \\
    \textbf{LP Filter} & 3rd order Butterworth \\
    \textbf{LP Filter BW (OOK-1/4)} & 4.32 MHz \\
    \textbf{LP Filter BW (OOK-2,FSK)} & $N_M$ SCs \\
    \textbf{WUS Sequence} $\mathbf{A}$ or $\mathbf{a}$ & Zadoff-Chu: $\mathbf{A}=\exp\{-j\pi u \frac{n(n+1)}{N_{ZC}}\}$ \\
    \textbf{WUS Sequence Parameters} & $u=1$, $n=0,1,...,143$, $N_{ZC}=149$ \\
    \textbf{Receiver} & Energy Detector \\
    \textbf{Channel Model} & AWGN, TDL-C 300ns \\
    \textbf{Impairments} & None \\
    \bottomrule
  \end{tabular}
  \vspace{5pt}
  \caption{Link-level simulation assumption \cite{lls_assumptions} agreed in email discussion [\texttt{Post-RAN1\#112-LP\_WUS2}]".}
  \label{tab:sim_assumptions}
\end{table}

% \vspace{-25pt}
\subsection{Comparison for $M=2$}
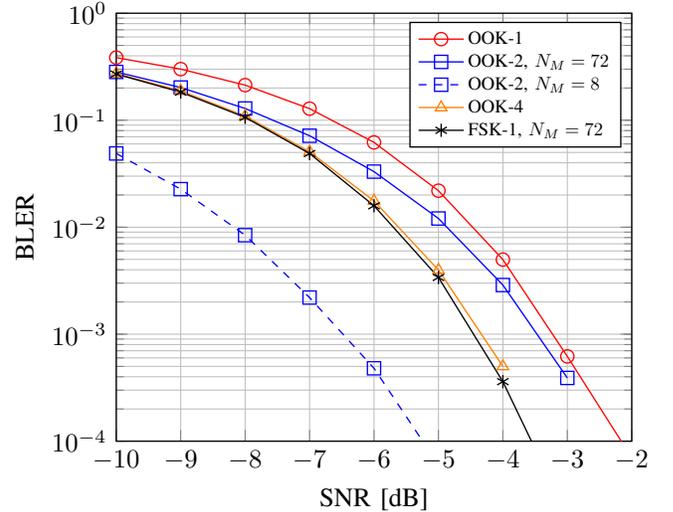
\begin{figure}[t]
  \centering
  \begin{tikzpicture}
  \tikzstyle{every pin}=[fill=white,draw=black]
   \pgfplotsset{every axis legend/.append style={
     cells={anchor=west},nodes={scale=0.7, transform shape}}}%, at={(0.5,1.05)}, anchor=south}}
 %   \pgfplotsset{every axis plot/.append style={smooth}}
    \pgfplotsset{every axis/.append style={line width=0.5pt}}
    \pgfplotsset{every axis/.append style={mark options=solid, mark size=2.5pt}}

    \begin{semilogyaxis}[xlabel={SNR [dB]}, ylabel={BLER},
      grid=minor, xmin=-10, xmax=-2, xtick={-10,-9,...,-2}, xmajorgrids, ymin=0.0001,
      ymax=1,ytickten={-4,-3,-2,-1,0}, legend columns=1]

      % OOK-1
      \addplot[red, solid, mark=o] plot coordinates {(-10.000000,0.384710) (-9.000000,0.300490) (-8.000000,0.212870) (-7.000000,0.128250) (-6.000000,0.062060) (-5.000000,0.021940) (-4.000000,0.004980) (-3.000000,0.000620) (-2.000000,0.000070) (-1.000000,0.000000) (0.000000,0.000000)};

      % OOK-2, BW=72, M=2, 2 symbols
      \addplot[blue, solid, mark=square] plot coordinates {(-10.000000,0.282550) (-9.000000,0.202100) (-8.000000,0.128680) (-7.000000,0.071600) (-6.000000,0.033080) (-5.000000,0.012060) (-4.000000,0.002880) (-3.000000,0.000390) (-2.000000,0.000000) (-1.000000,0.000000) (0.000000,0.000000)};

      % OOK-2, BW=48, M=2, 2 symbols
      % \addplot[blue, dashed, mark=square] plot coordinates {(-10.000000,0.222550) (-9.000000,0.151600) (-8.000000,0.093900) (-7.000000,0.049940) (-6.000000,0.022880) (-5.000000,0.007330) (-4.000000,0.001620) (-3.000000,0.000180) (-2.000000,0.000030) (-1.000000,0.000010) (0.000000,0.000000)};

      % OOK-2, BW=8, M=2, 2 symbols
      \addplot[blue, dashed, mark=square] plot coordinates {(-10.000000,0.048900) (-9.000000,0.022680) (-8.000000,0.008420) (-7.000000,0.002200) (-6.000000,0.000480) (-5.000000,0.000060) (-4.000000,0.000000) (-3.000000,0.000000) (-2.000000,0.000000) (-1.000000,0.000000) (0.000000,0.000000)};

      % OOK-4, M=2, 2 symbols
      \addplot[orange, solid, mark=triangle] plot coordinates {(-10.000000,0.272940) (-9.000000,0.188070) (-8.000000,0.109450) (-7.000000,0.050730) (-6.000000,0.017610) (-5.000000,0.003980) (-4.000000,0.000500) (-3.000000,0.000000) (-2.000000,0.000000) (-1.000000,0.000000) (0.000000,0.000000)};

      % FSK-1, BW=72, M=2, 2 symbols
      \addplot[black, solid, mark=asterisk] plot coordinates {(-10.000000,0.270000) (-9.000000,0.183140) (-8.000000,0.106420) (-7.000000,0.048840) (-6.000000,0.015860) (-5.000000,0.003390) (-4.000000,0.000360) (-3.000000,0.000020) (-2.000000,0.000000) (-1.000000,0.000000) (0.000000,0.000000)};

      \legend{ {OOK-1}\\
               {OOK-2, $N_M=72$}\\
               {OOK-2, $N_M=8$}\\
               {OOK-4}\\
               {FSK-1, $N_M=72$}\\};

    \end{semilogyaxis}
  \end{tikzpicture}
  \vspace{-10pt}
  \caption{Performance comparison of various LP-WUS waveforms, AWGN, $B=2$, $M=2$, $L=2$, OOK-1: $M=1$, $L=4$.}
  \label{fig:awgn_m2}
\end{figure}

The results in Figure \ref{fig:awgn_m2} show the performance in AWGN for a payload of $B=2$ bits and Manchester coding. Consequently, OOK-1 requires $L=4$ symbols of which only 2 are active for any payload. OOK-2 and FSK-1 are configured \textit{without} inner GB ($N_M=72$), for a fair comparison with OOK-4, because a GB would reduce the segment size $N_M$ and hence the LP filter BW resulting in an additional SNR gain since less noise is captured. Additionally, we apply a power boost of $10\log_{10}(4/3)$ dB for OOK-2 so that the total average transmit power is the same as for the other schemes. This is necessary since for payloads $\mathbf{b}=[0, 0]$ and $\mathbf{b}=[1, 1]$ OOK-2 is only transmitting in one OFDM symbol because of Manchester coding in time domain.

From Figure \ref{fig:awgn_m2}, we observe that OOK-4 is outperforming OOK-1 by $~1.2$dB at 1\%BLER which is due to the fact that all the transmit power is concentrated in a shorter OOK symbol, the OOK-4 symbol is half as long as OOK-1, resulting in a higher SNR per OOK symbol. OOK-2 performs worse than OOK-4 because of interference between OOK symbols since OOK-2 applies Manchester coding over time domain. More precisely, the imperfect LPF will capture energy from the neighboring segment. For payloads $\mathbf{b}=[1, 0]$ and $\mathbf{b}=[0, 1]$ this interference impacts the detection performance. The performance of OOK-2 can be improved by using a GB ($N_M=8$) and therefore shortening the segments. This reduces the BW of the LPF which captures less noise and thus the SNR per segment (OOK symbol) increases. Note that in the extreme case, only a single SC may be active. However, as we will see later, the performance in frequency-selective channels suffers significantly.

FSK-1 has essentially the same performance as OOK-4. Moreover, for $M=2$, FSK-1 does not suffer from interference of neighboring segments as OOK-2 because Manchester coding is applied in frequency-domain and hence only a single segment is active per OFDM symbol.

\begin{figure}[t]
  \centering
  \begin{tikzpicture}
  \tikzstyle{every pin}=[fill=white,draw=black]
  \pgfplotsset{every axis legend/.append style={
     cells={anchor=west},nodes={scale=0.7, transform shape}}}%, at={(0.5,1.05)}, anchor=south}}
 %   \pgfplotsset{every axis plot/.append style={smooth}}
    \pgfplotsset{every axis/.append style={line width=0.5pt}}
    \pgfplotsset{every axis/.append style={mark options=solid, mark size=2.5pt}}

    \begin{semilogyaxis}[xlabel={SNR [dB]}, ylabel={BLER},
      grid=minor, xmin=-8, xmax=5, xtick={-8,-7,...,5}, xmajorgrids, ymin=0.001,
      ymax=1,ytickten={-3,-2,-1,0}, legend columns=1]

      % OOK-1
      \addplot[red, solid, mark=o] plot coordinates {(-10.000000,0.401580) (-9.000000,0.339400) (-8.000000,0.276470) (-7.000000,0.217190) (-6.000000,0.163820) (-5.000000,0.117600) (-4.000000,0.081670) (-3.000000,0.053730) (-2.000000,0.033200) (-1.000000,0.019260) (0.000000,0.010050) (1.000000,0.004950) (2.000000,0.002340) (3.000000,0.000920) (4.000000,0.000340) (5.000000,0.000120) (6.000000,0.000040) (7.000000,0.000000) (8.000000,0.000000) (9.000000,0.000000) (10.000000,0.000000)};

      % OOK-2, BW=72, M=2, 2 symbols
      \addplot[blue, solid, mark=square] plot coordinates {(-10.000000,0.363890) (-9.000000,0.309230) (-8.000000,0.257170) (-7.000000,0.207080) (-6.000000,0.161640) (-5.000000,0.122240) (-4.000000,0.089240) (-3.000000,0.062220) (-2.000000,0.041430) (-1.000000,0.026450) (0.000000,0.016010) (1.000000,0.009350) (2.000000,0.005260) (3.000000,0.002740) (4.000000,0.001470) (5.000000,0.000760) (6.000000,0.000400) (7.000000,0.000170) (8.000000,0.000080) (9.000000,0.000070) (10.000000,0.000060) };

      % OOK-2, BW=48, M=2, 2 symbols
      % \addplot[blue, dashed, mark=square] plot coordinates {(-10.000000,0.321000) (-9.000000,0.268600) (-8.000000,0.222800) (-7.000000,0.180500) (-6.000000,0.144900) (-5.000000,0.109200) (-4.000000,0.081400) (-3.000000,0.060500) (-2.000000,0.041300) (-1.000000,0.027500) (0.000000,0.016600) (1.000000,0.010400) (2.000000,0.005100) (3.000000,0.003200) (4.000000,0.001500) (5.000000,0.000700) (6.000000,0.000400) (7.000000,0.000300) (8.000000,0.000100) (9.000000,0.000100) (10.000000,0.000000)};

      % OOK-2, BW=8, M=2, 2 symbols
      \addplot[blue, dashed, mark=square] plot coordinates {(-10.000000,0.213930) (-9.000000,0.177730) (-8.000000,0.146310) (-7.000000,0.118530) (-6.000000,0.094540) (-5.000000,0.074730) (-4.000000,0.058430) (-3.000000,0.045210) (-2.000000,0.034150) (-1.000000,0.025560) (0.000000,0.019310) (1.000000,0.014350) (2.000000,0.010590) (3.000000,0.007610) (4.000000,0.005430) (5.000000,0.003650) (6.000000,0.002350) (7.000000,0.001520) (8.000000,0.001110) (9.000000,0.000740) (10.000000,0.000470) };

      % OOK-4, M=2, 2 symbols
      \addplot[orange, solid, mark=triangle] plot coordinates {(-10.000000,0.325590) (-9.000000,0.264530) (-8.000000,0.208070) (-7.000000,0.156650) (-6.000000,0.112890) (-5.000000,0.078270) (-4.000000,0.052350) (-3.000000,0.033280) (-2.000000,0.019730) (-1.000000,0.010770) (0.000000,0.005700) (1.000000,0.002950) (2.000000,0.001370) (3.000000,0.000690) (4.000000,0.000320) (5.000000,0.000120) (6.000000,0.000030) (7.000000,0.000000) (8.000000,0.000000) (9.000000,0.000000) (10.000000,0.000000) };

      % FSK-1, BW=72, M=2, 2 symbols
      \addplot[black, solid, mark=asterisk] plot coordinates {(-10.000000,0.348670) (-9.000000,0.294480) (-8.000000,0.242610) (-7.000000,0.192450) (-6.000000,0.147490) (-5.000000,0.109920) (-4.000000,0.078220) (-3.000000,0.053400) (-2.000000,0.034680) (-1.000000,0.021760) (0.000000,0.013160) (1.000000,0.007240) (2.000000,0.004100) (3.000000,0.002160) (4.000000,0.000970) (5.000000,0.000430) (6.000000,0.000180) (7.000000,0.000080) (8.000000,0.000040) (9.000000,0.000010) (10.000000,0.000010)};

      \legend{ {OOK-1}\\
               {OOK-2, $N_M=72$}\\
               {OOK-2, $N_M=8$}\\
               {OOK-4}\\
               {FSK-1, $N_M=72$}\\};

    \end{semilogyaxis}
  \end{tikzpicture}
  \vspace{-10pt}
  \caption{Performance comparison of various LP-WUS waveforms, TDL-C 300ns, $B=2$, $M=2$, $L=2$, OOK-1: $M=1$, $L=4$.}
  \label{fig:tdlc_m2}
\end{figure}
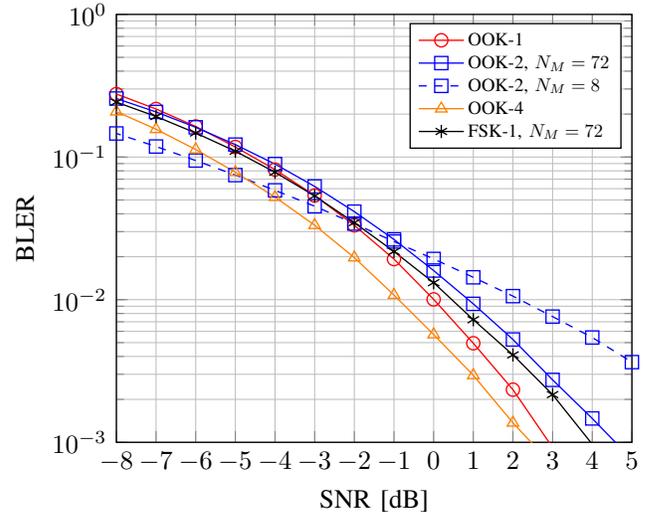

Figure \ref{fig:tdlc_m2} compares the same schemes in a frequency-selective channel. It can be observed that the techniques dividing the BW into segments, i.e. OOK-2 and FSK, suffer a performance loss compared to OOK-1/4. The reason is the loss of frequency diversity. OOK-2 with $N_M=8$ illustrates this well, the small BW makes it vulnerable to deep fades in the channel frequency response. At low SNR, where the noise is dominating, OOK-2 outperforms the other schemes since the LPF rejects noise well. 

\vspace{-10pt}
\subsection{Comparison for $M=4$}

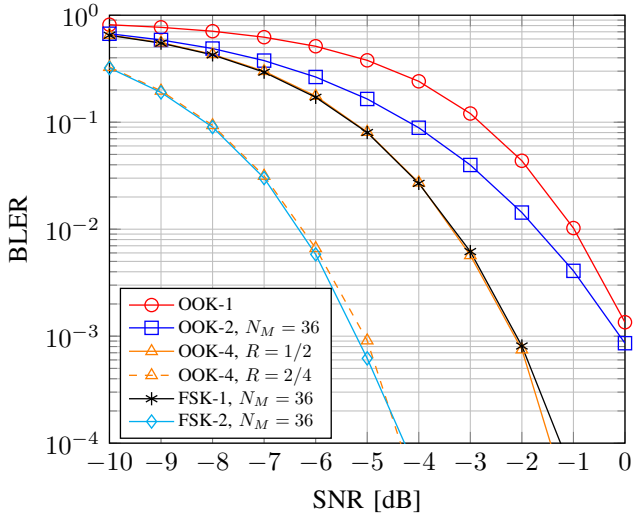
\begin{figure}[t]
  \centering
  \begin{tikzpicture}
  \tikzstyle{every pin}=[fill=white,draw=black]
   \pgfplotsset{every axis legend/.append style={
     cells={anchor=west},nodes={scale=0.7, transform shape}, at={(.02,0.01)}, anchor=south west}}
 %   \pgfplotsset{every axis plot/.append style={smooth}}
    \pgfplotsset{every axis/.append style={line width=0.5pt}}
    \pgfplotsset{every axis/.append style={mark options=solid, mark size=2.5pt}}

    \begin{semilogyaxis}[xlabel={SNR [dB]}, ylabel={BLER},
      grid=minor, xmin=-10, xmax=0, xtick={-10,-9,...,0}, xmajorgrids, ymin=0.0001,
      ymax=1,ytickten={-4,-3,-2,-1,0}, legend columns=1]

      % OOK-1
      \addplot[red, solid, mark=o] plot coordinates {(-10.000000,0.815130) (-9.000000,0.770750) (-8.000000,0.708840) (-7.000000,0.623240) (-6.000000,0.512750) (-5.000000,0.378870) (-4.000000,0.241140) (-3.000000,0.120590) (-2.000000,0.043600) (-1.000000,0.010250) (0.000000,0.001350)};

      % OOK-2, BW=36, M=4, 2 symbols
      \addplot[blue, solid, mark=square] plot coordinates {(-10.000000,0.671340) (-9.000000,0.587860) (-8.000000,0.488290) (-7.000000,0.376710) (-6.000000,0.264660) (-5.000000,0.164910) (-4.000000,0.088770) (-3.000000,0.039840) (-2.000000,0.014310) (-1.000000,0.004080) (0.000000,0.000860)};

      % OOK-4, M=4, 2 symbols, R=1/2
      \addplot[orange, solid, mark=triangle] plot coordinates {(-10.000000,0.653860) (-9.000000,0.553850) (-8.000000,0.431430) (-7.000000,0.299660) (-6.000000,0.175760) (-5.000000,0.080930) (-4.000000,0.027240) (-3.000000,0.005700) (-2.000000,0.000750) (-1.000000,0.000020) (0.000000,0.000010)};

      % OOK-4, M=4, 2 symbols, R=2/4
      \addplot[orange, dashed, mark=triangle] plot coordinates {(-10.000000,0.327230) (-9.000000,0.196490) (-8.000000,0.093600) (-7.000000,0.031460) (-6.000000,0.006640) (-5.000000,0.000910) (-4.000000,0.000030) (-3.000000,0.000000) (-2.000000,0.000000) (-1.000000,0.000000) (0.000000,0.000000)};

      % FSK-1, BW=36, M=4, 2 symbols
      \addplot[black, solid, mark=asterisk] plot coordinates {(-10.000000,0.648710) (-9.000000,0.548420) (-8.000000,0.426220) (-7.000000,0.293400) (-6.000000,0.170840) (-5.000000,0.080080) (-4.000000,0.026810) (-3.000000,0.006180) (-2.000000,0.000810) (-1.000000,0.000050) (0.000000,0.000000)};

       % FSK-2, BW=36, M=4, 2 symbols
      \addplot[cyan, solid, mark=diamond] plot coordinates {(-10.000000,0.321230) (-9.000000,0.189960) (-8.000000,0.090040) (-7.000000,0.030040) (-6.000000,0.005820) (-5.000000,0.000620) (-4.000000,0.000050) (-3.000000,0.000000) (-2.000000,0.000000) (-1.000000,0.000000) (0.000000,0.000000)};

      \legend{ {OOK-1}\\
               {OOK-2, $N_M=36$}\\
               {OOK-4, $R=1/2$}\\
               {OOK-4, $R=2/4$}\\
               {FSK-1, $N_M=36$}\\
               {FSK-2, $N_M=36$}\\};

    \end{semilogyaxis}
  \end{tikzpicture}
  \vspace{-10pt}
  \caption{Performance comparison of various LP-WUS waveforms, AWGN, $B=4$, $M=4$, $L=2$, OOK-1: $M=1$, $L=8$.}
  \label{fig:awgn_m4}
\end{figure}

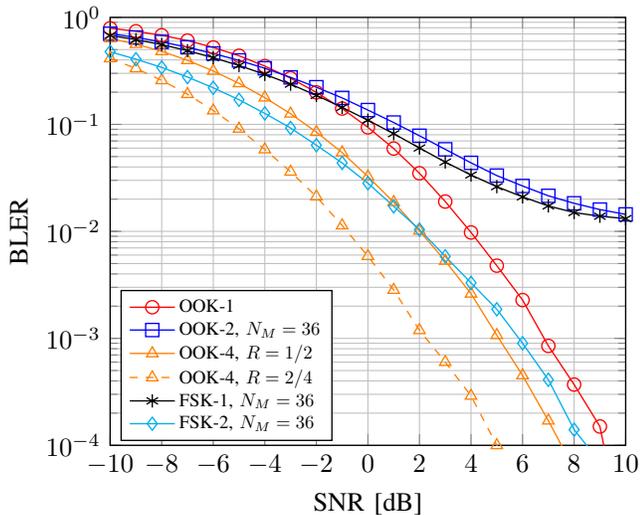
\begin{figure}[t]
  \centering
  \begin{tikzpicture}
  \tikzstyle{every pin}=[fill=white,draw=black]
   \pgfplotsset{every axis legend/.append style={
     cells={anchor=west},nodes={scale=0.7, transform shape}, at={(.02,0.01)}, anchor=south west}}
 %   \pgfplotsset{every axis plot/.append style={smooth}}
    \pgfplotsset{every axis/.append style={line width=0.5pt}}
    \pgfplotsset{every axis/.append style={mark options=solid, mark size=2.5pt}}

    \begin{semilogyaxis}[xlabel={SNR [dB]}, ylabel={BLER},
      grid=minor, xmin=-10, xmax=10, xtick={-10,-8,...,10}, xmajorgrids, ymin=0.0001,
      ymax=1,ytickten={-4,-3,-2,-1,0}, legend columns=1]

      % OOK-1
      \addplot[red, solid, mark=o] plot coordinates {(-10.000000,0.791840) (-9.000000,0.741800) (-8.000000,0.680400) (-7.000000,0.607200) (-6.000000,0.523800) (-5.000000,0.437870) (-4.000000,0.350650) (-3.000000,0.269990) (-2.000000,0.199210) (-1.000000,0.140480) (0.000000,0.093770) (1.000000,0.059450) (2.000000,0.034920) (3.000000,0.019000) (4.000000,0.009770) (5.000000,0.004780) (6.000000,0.002280) (7.000000,0.000850) (8.000000,0.000370) (9.000000,0.000150) (10.000000,0.000010)};

      % OOK-2, BW=36, M=4, 2 symbols
      \addplot[blue, solid, mark=square] plot coordinates {(-10.000000,0.703400) (-9.000000,0.649630) (-8.000000,0.589760) (-7.000000,0.526020) (-6.000000,0.459850) (-5.000000,0.396280) (-4.000000,0.334580) (-3.000000,0.275230) (-2.000000,0.222610) (-1.000000,0.176730) (0.000000,0.136080) (1.000000,0.104040) (2.000000,0.078250) (3.000000,0.058470) (4.000000,0.043700) (5.000000,0.033220) (6.000000,0.026560) (7.000000,0.021490) (8.000000,0.018290) (9.000000,0.015980) (10.000000,0.014410)};

      % OOK-4, M=4, 2 symbols, R=1/2
      \addplot[orange, solid, mark=triangle] plot coordinates {(-10.000000,0.636360) (-9.000000,0.562310) (-8.000000,0.481620) (-7.000000,0.397490) (-6.000000,0.315530) (-5.000000,0.241750) (-4.000000,0.177820) (-3.000000,0.125760) (-2.000000,0.084710) (-1.000000,0.054480) (0.000000,0.032500) (1.000000,0.018740) (2.000000,0.010110) (3.000000,0.005290) (4.000000,0.002610) (5.000000,0.001070) (6.000000,0.000450) (7.000000,0.000170) (8.000000,0.000060) (9.000000,0.000030) (10.000000,0.000010)};

      % OOK-4, M=4, 2 symbols, R=2/4
      \addplot[orange, dashed, mark=triangle] plot coordinates {(-10.000000,0.413470) (-9.000000,0.333240) (-8.000000,0.257950) (-7.000000,0.191460) (-6.000000,0.134620) (-5.000000,0.090820) (-4.000000,0.058270) (-3.000000,0.036000) (-2.000000,0.021140) (-1.000000,0.011340) (0.000000,0.005860) (1.000000,0.002830) (2.000000,0.001190) (3.000000,0.000600) (4.000000,0.000290) (5.000000,0.000100) (6.000000,0.000030) (7.000000,0.000000) (8.000000,0.000000) (9.000000,0.000000) (10.000000,0.000000)};

      % FSK-1, BW=36, M=4, 2 symbols
      \addplot[black, solid, mark=asterisk] plot coordinates {(-10.000000,0.677810) (-9.000000,0.619260) (-8.000000,0.557270) (-7.000000,0.488490) (-6.000000,0.420350) (-5.000000,0.355070) (-4.000000,0.293520) (-3.000000,0.237550) (-2.000000,0.186560) (-1.000000,0.144090) (0.000000,0.109350) (1.000000,0.081810) (2.000000,0.060150) (3.000000,0.044440) (4.000000,0.033630) (5.000000,0.026160) (6.000000,0.021010) (7.000000,0.017280) (8.000000,0.014990) (9.000000,0.013850) (10.000000,0.013160)};

       % FSK-2, BW=36, M=4, 2 symbols
      \addplot[cyan, solid, mark=diamond] plot coordinates {(-10.000000,0.477030) (-9.000000,0.406850) (-8.000000,0.340550) (-7.000000,0.277000) (-6.000000,0.219730) (-5.000000,0.169880) (-4.000000,0.126600) (-3.000000,0.092480) (-2.000000,0.063890) (-1.000000,0.043610) (0.000000,0.028180) (1.000000,0.017140) (2.000000,0.010390) (3.000000,0.005840) (4.000000,0.003300) (5.000000,0.001870) (6.000000,0.000900) (7.000000,0.000410) (8.000000,0.000140) (9.000000,0.000070) (10.000000,0.000020)};

      \legend{ {OOK-1}\\
               {OOK-2, $N_M=36$}\\
               {OOK-4, $R=1/2$}\\
               {OOK-4, $R=2/4$}\\
               {FSK-1, $N_M=36$}\\
               {FSK-2, $N_M=36$}\\};

    \end{semilogyaxis}
  \end{tikzpicture}
  \vspace{-10pt}
  \caption{Performance comparison of various LP-WUS waveforms, TDL-C, $B=4$, $M=4$, $L=2$, OOK-1: $M=1$, $L=8$.}
  \label{fig:tdlc_m4}
\end{figure}

Figures \ref{fig:awgn_m4} and \ref{fig:tdlc_m4} compare the performance for $M=4$ in AWGN and TDL-C, respectively. For OOK-2, we apply a power boost of $10\log_{10}(32/30)$ dB to ensure the same average transmit power. It can be observed that in AWGN, "OOK-4, R=1/2" performs similar to FSK-1. Moreover, if the same encoding scheme as for FSK-2 is applied to OOK-4, i.e., "OOK-4, R=2/4", both schemes have identical performance. On the other hand, in the TDL-C channel, all schemes partitioning the WUS bandwidth perform significantly worse than the schemes that utilise the entire WUS BW. For instance, there is a performance difference of almost 3dB at 1\% BLER between FSK-2 and "OOK-4, R=2/4".

\subsection{Time-Domain Overlay Code}

\begin{figure}[t]
  \centering
  \begin{tikzpicture}
  \tikzstyle{every pin}=[fill=white,draw=black]
  \pgfplotsset{every axis legend/.append style={
     cells={anchor=west},nodes={scale=0.7, transform shape}}}
  % \pgfplotsset{every axis legend/.append style={
     % cells={anchor=west},nodes={scale=0.7, transform shape}, at={(0.5,1.05)}, anchor=south}}
 %   \pgfplotsset{every axis plot/.append style={smooth}}
    \pgfplotsset{every axis/.append style={line width=0.5pt}}
    \pgfplotsset{every axis/.append style={mark options=solid, mark size=2.5pt}}

    \begin{semilogyaxis}[xlabel={SNR [dB]}, ylabel={BLER},
      grid=minor, xmin=-16, xmax=12, xtick={-16,-12,...,12}, xmajorgrids, ymin=0.001,
      ymax=1,ytickten={-3,-2,-1,0}, legend columns=1]

      % OOK-4, M=8, Rate=1/2, R=7
      \addplot[red, solid, mark=o] plot coordinates {(-12, 0.7134) (-10, 0.5233) (-8, 0.3213) (-6, 0.1711) (-4, 0.0821) (-2, 0.0367) (0, 0.013) (2, 0.0023) (4, 0.0003)};
      
      % OOK-4, M=8, Rate=2/4, R=7
      \addplot[red, dashed, mark=o] plot coordinates {(-16, 0.8021) (-14, 0.6393) (-12, 0.4362) (-10, 0.2491) (-8, 0.1227) (-6, 0.0581) (-4, 0.0243) (-2, 0.0057) (0, 0.0006)};

      % OOK-4, M=8, Rate=3/8, R=14, Bh=5 bits, overall 
      \addplot[blue, solid, mark=square] plot coordinates {(-16, 0.27953) (-14, 0.15918) (-12, 0.07935) (-10, 0.03558) (-8, 0.01587) (-6, 0.00752) (-4, 0.00418) (-2, 0.00261)};

      % OOK-4, M=8, Rate=3/8, R=14, Bh=5 bits, WUS bits 
      \addplot[blue, dashed, mark=square] plot coordinates {(-16, 0.25295) (-14, 0.13854) (-12, 0.06354) (-10, 0.02369) (-8, 0.00709) (-6, 0.0017) (-4, 0.00022) (-2, 3.00E-05)};

      % OOK-4, M=2, Rate=1/2, R=14, Bh=7 bits, overall 
      \addplot[orange, solid, mark=triangle] plot coordinates {(-16, 0.2427) (-14, 0.1478) (-12, 0.079) (-10, 0.0425) (-8, 0.0233) (-6, 0.0119) (-4, 0.0065) (-2, 0.0041) };

      % OOK-4, M=2, Rate=1/2, R=14, Bh=7 bits, WUS bits 
      \addplot[orange, dashed, mark=triangle] plot coordinates {(-16, 0.1768) (-14, 0.1074) (-12, 0.0535) (-10, 0.0234) (-8, 0.0096) (-6, 0.0031)};

      \legend{ {$M=8$, $R=1/2$, $L=2$, $L_R=7$}\\
               {$M=8$, $R=2/4$, $L=2$, $L_R=7$}\\
               {$M=8$, $R=3/8$, $L_R=14$, $B_h=5$}\\
               %{$M=8$, $R=3/8$, $L_RR=14$, $B_h=5$}\\
               {$M=2$, $R=1/2$, $L_R=14$, $B_h=7$}\\};
               %{$M=2$, $R=1/2$, $L_R=14$, $B_h=7$}\\};

    \draw (axis cs:6,0.02)  node[fill=white,draw=black] (pint0) {overall}; 
    \draw (axis cs:-6,0.01) node[draw,black,ellipse,minimum width=0.5cm] (ell0) {}; 
    \draw[black] (pint0) -- (ell0);

    \draw (axis cs:-12,0.003)  node[fill=white,draw=black] (pint1) {$B_v$ bits}; 
    \draw (axis cs:-8.1,0.01) node[draw,black,ellipse,minimum width=0.3cm] (ell1) {}; 
    \draw[black] (pint1) -- (ell1);

    \end{semilogyaxis}
  \end{tikzpicture}
  \vspace{-10pt}
  \caption{Performance comparison of various OOK-4 waveforms with overlay code, TDL-C 300ns, 14 OFDM symbols, $B=8$.}
  \label{fig:tdlc_vh}
\end{figure}
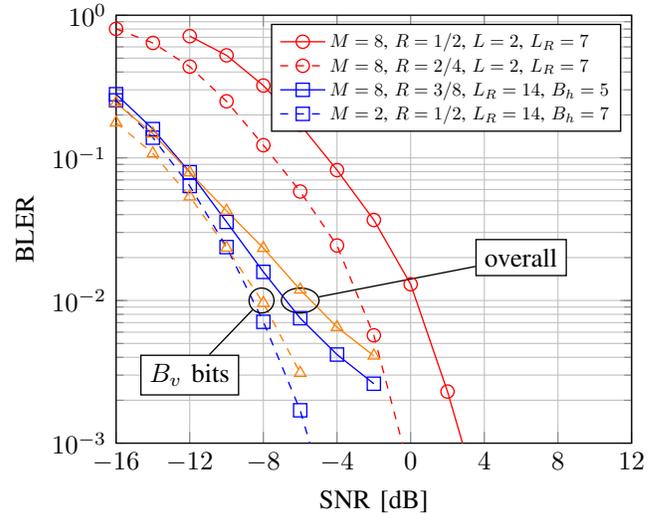

In Figure \ref{fig:tdlc_vh}, we compare the performance of OOK-4 with and without an overlay code for a payload of $B=8$ bits and 14 OFDM symbols, i.e. one slot. The reference scheme (red lines) uses $L=2$ consecutive OFDM symbols and a repetition factor of $L_R=7$. The overlay code of $B_h=5$ (blue lines) and $B_h=7$ (orange lines) is applied to a OOK-4 modulation with $M=8$ and $M=2$, respectively. For $B_h=5$ we use a Manchester code with rate $R=3/8$ to transmit $B_v=3$ bits. 
It can be observed that the schemes with overlay code significantly outperforms the reference schemes at the expense of moderately increases receiver complexity. For $M=8$ and $M=2$ the gain is almost 5dB and 4dB at 1\%BLER, respectively. Moreover, in the schemes with overlay code, the $B_v$ bits have lower BLER and benefit from increased error protection. Consequently, decreasing $M$ (i.e. decreasing $B_v$) and increasing $B_h$ reduces the overall performance. The simulation results show that spectral efficiency is significantly increased (i.e. more bits can be transmitted over the same BW) while only requiring a slightly higher SNR at the receiver.

\section{Conclusion}\label{sec:conclusion}

In this paper, we present a concise summary of LP-WUS waveforms discussed in the 3GPP Rel-18 study item and provide simulation results of their respective performance in different propagation scenarios. Moreover, we propose a time-domain overlay coding scheme and show that it can significantly increase spectral efficiency with moderate receiver complexity. Low-power communication steers a lot of interest within the 3GPP community because of its vast market opportunities. Many companies support the standardization of LP-WUS in Rel-19 as well as the study of the related subject called "Ambient-IoT" investigating the integration of ultra-low power devices into the 5G ecosystem.

\section*{Acknowledgment}
This work has been supported by Qualcomm.

% The preferred spelling of the word ``acknowledgment'' in America is without
% an ``e'' after the ``g''. Avoid the stilted expression ``one of us (R. B.
% G.) thanks $\ldots$''. Instead, try ``R. B. G. thanks$\ldots$''. Put sponsor
% acknowledgments in the unnumbered footnote on the first page.

%     \bibitem{b1} G. Eason, B. Noble, and I. N. Sneddon, ``On certain integrals of Lipschitz-Hankel type involving products of Bessel functions,'' Phil. Trans. Roy. Soc. London, vol. A247, pp. 529--551, April 1955.
%     \bibitem{b2} J. Clerk Maxwell, A Treatise on Electricity and Magnetism, 3rd ed., vol. 2. Oxford: Clarendon, 1892, pp.68--73.
%     \bibitem{b3} I. S. Jacobs and C. P. Bean, ``Fine particles, thin films and exchange anisotropy,'' in Magnetism, vol. III, G. T. Rado and H. Suhl, Eds. New York: Academic, 1963, pp. 271--350.
%     \bibitem{b4} K. Elissa, ``Title of paper if known,'' unpublished.
%     \bibitem{b5} R. Nicole, ``Title of paper with only first word capitalized,'' J. Name Stand. Abbrev., in press.
%     \bibitem{b6} Y. Yorozu, M. Hirano, K. Oka, and Y. Tagawa, ``Electron spectroscopy studies on magneto-optical media and plastic substrate interface,'' IEEE Transl. J. Magn. Japan, vol. 2, pp. 740--741, August 1987 [Digests 9th Annual Conf. Magnetics Japan, p. 301, 1982].
%     \bibitem{b7} M. Young, The Technical Writer's Handbook. Mill Valley, CA: University Science, 1989.
% \end{thebibliography}
\vspace{12pt}

\end{document}